# Energy Efficiency of Fog Computing Health Monitoring Applications

Ida Syafiza M. Isa, Mohamed O.I. Musa, Taisir E.H. El-Gorashi, Ahmed Q. Lawey and Jaafar M. H. Elmirghani

*School of Electronic and Electrical Engineering, University of Leeds, Leeds, LS2 9JT, UK*

**ABSTRACT**

Fog computing offers a scalable and effective solution to overcome the increasing processing and networking demands of Internet of Thing (IoT) devices. In this paper, we investigate the use of fog computing for health monitoring applications. We consider a heart monitoring application where patients send their 30 minute recording of Electrocardiogram (ECG) signal for processing, analysis, and decision making at fog processing units within the time constraint recommended by the American Heart Association (AHA) to save heart patients when an abnormality in the ECG signal is detected. The locations of the processing servers are optimized so that the energy consumption of both the processing and networking equipment are minimised. The results show that processing the ECG signal at fog processing units yields a total energy consumption saving of up to 68% compared to processing the at the central cloud.

## 1. INTRODUCTION

The recent increase in chronic diseases and the aging population have triggered the revolution of remote health monitoring in developed countries. The advancement of wireless body sensors together with the growing cloud computing applications that offer high processing and storage capabilities for health data has supported the development of health monitoring systems that provide real-time response to the patients pertaining to their health condition. However, the transmission of health-related data from a large number of patients to the cloud contributes to increase the congestion in cloud networking infrastructures which leads to high latency and hence violations to Quality-of-Service (QoS) metrics [1]. Fog Computing offers a viable solution that can address the challenges related to the delivery of QoS in healthcare monitoring system to patients due to reduced latency. Furthermore, fog computing reduce the energy consumption in cloud networking infrastructures [2]–[10] under increasing applications traffic.

Several studies have investigated the use of fog computing to develop efficient health monitoring systems. For instance, a monitoring system in [1] employed the concept of fog computing at a smart gateway to efficiently process health data, particularly Electrocardiogram (ECG) signals. The ECG empirical results for feature extraction using the proposed system achieved 90% bandwidth efficiency and real-time response with low latency. Azimi et al. [11] claimed that both continuous monitoring and real-time monitoring may not be feasible with the current IoT-based systems to perform data analysis. Hence, fog computing was embedded into the system, which results in reduced response time and increased system reliability when the Internet connection is not available. A prototype of a smart e-health gateway (i.e. the fog computing device) is implemented in [12]. It performs high level services such as real-time data processing, local storage, and embedded data mining to reduce the burden at the sensor node and the cloud. The performance of the gateway is evaluated in terms of energy efficiency of the sensor nodes, scalability, mobility and reliability. Meanwhile, a real-time event triggering for health monitoring systems in smart homes is proposed in [13] by implementing the Bayesian Belief Network (BBN) to classify the state of an event at the fog layer.

In this work, we optimise the placement of processing servers at the network edge so that the total energy consumption of the ECG health monitoring application is minimised. The rest of this paper is organized as follows: Section 2 describes the fog computing architecture to serve the ECG monitoring application. Section 3 presents and analyses the results and Section 4 gives the conclusions.

## 2. FOG COMPUTING ARCHITECTURE FOR HEALTH MONITORING APPLICATIONS

Figure 1 shows the end-to-end network connecting patients to the central cloud. The network consists of three layers: the access network, metro network and core network. Fog computing processing resources serving the health monitoring application can be deployed in two layers in the access network. The first layer consists of processing servers (PS) connected to the Optical Network Terminals (ONT) of the Gigabit Passive Optical Network (GPON) [14]. Placing the processing servers in this layer, which is closer to the users, decreases the energy consumption of networking equipment, however, it will increase the required number of processing servers. The second layer consists of processing servers connected to the Optical Line Terminal (OLT). Utilising processing servers in this layer reduces the number of required processing servers as it is a shared point between the access points but will increase the energy consumption of the networking equipment. We developed a MILP model to optimise serving the health monitoring application patients from these two layers, so the total energy consumption is minimised.

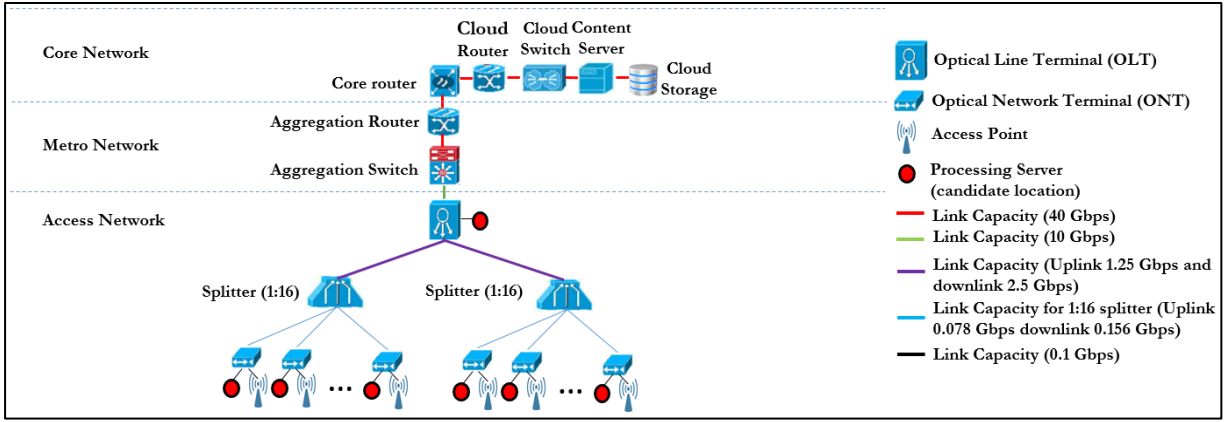

*Figure 1. GPON architecture with fog network*

## 3. RESULTS

In this section we optimise serving the ECG monitoring application patients in the fog architecture described above considering a scenario with 200 patients uniformly distributed among 32 Wi-Fi access points. Each access point is connected to an ONT to aggregate the ECG signal from patients. All the ONTs are connected to a single OLT. Each patient transmits their own 30 minutes ECG recording signal with a size of 1.92 Mbits [1] to the network to be processed and analysed at the fog. The processed ECG signal, which is 126.72 kbits [1], is to be permanently stored at the cloud storage.

The overall energy consumption calculated in this work is based on the timing constraints set by health monitoring standards. According to the American Heart Association (AHA), the golden time to save a heart patient by sending an alarm message to a cardiologist when detecting a sudden fall or rise in cardiac vital signs is within 4 to 6 minutes [15]. Accordingly, $T_T$=4 minutes for has been selected in this work as the maximum duration imposed by AHA. This duration together with the data volume to be transmitted are used to calculate the minimum data rate needed for each patient. Note that the 4 minutes include the transmission time to send the ECG signal to the processing server, $T_t$, and the time to perform the processing and analysis, $T_{pa}$ (i.e. $T_T = T_t + T_{pa}$). The time required to process the 30 minutes ECG signal for each patient is 96.3 ms [1]. We assume that the analysis time is 10% of the processing time, hence the total time of both processing and analysis per patient is 105.9 ms. To determine the available time to transmit the ECG signal to the processing server, $T_t$, we calculate the time of both processing and analysis, $T_{pa}$ based on the maximum number of patients that can be served by a processing server ($Pat_{max}$), (i.e. $T_{pa} = 105.9\ ms \times Pat_{max}$) for $T_T$ equals 4 minutes. We investigate four scenarios where $Pat_{max}$ values are 50 patients (scenario 1), 100 patients (scenario 2), 150 patients (scenario 3) and 200 patients (scenario 4). For each user, the minimum data rate needed to transmit the 30 minutes ECG signal in $T_t$ seconds is 1.92 $Mbits/T_t$. Note that the higher the $Pat_{max}$, the higher the data rate given to each patient to transmit ECG signal to the processing sever.

The minimum data rate to send the analysed ECG data for permanent storage at the central cloud is calculated as the follows: We considered a scenario where the processing server is assigned to serve $Pat_{max}$, we then determine the minimum uplink capacity between the candidate locations of processing server to the cloud storage (i.e link between ONT and OLT). As the uplink capacity will be shared by the maximum number of patients, the processing server can serve, we divide the uplink capacity by $Pat_{max}$ to obtain the data rate for each patient to transmit the analysed ECG signal to the cloud storage. Note that, the uplink capacities are shared by multiple applications. Machine-to-machine (M2M) traffic is typically 5% of the global traffic [16] and 6% of the M2M traffic is attributed to healthcare applications [17]. Hence, in this work, we consider 0.3% of the maximum available link capacity is dedicated for healthcare applications. Table 1 shows the maximum time for processing and analysis, data rate and transmission time to transmit the ECG signal to the processing server, and the data rate and transmission time to transmit the analysed ECG signal to cloud storage in the four scenarios.

In each $Pat_{max}$ scenario, we proposed 2 approaches for the optimisation: Single processing server in Fog Approach (SFA) where only one processing server can be located at each location, and Multiple processing servers in Fog Approach (MFA) where more than one processing server can be located at the OLT while only one at each ONT. The two approaches are used to investigate the impact of limiting the number of processing servers at the OLT to 1 in SFA compared to allowing multiple processing servers at the OLT in the MFA. The impact is observed in terms of the distribution of the remaining required processing servers to serve $Pat_{max}$ in each scenario while minimising the energy consumption of the network and processing.

*Table 1. Data rates and related times for the different scenarios.*

| Type of Scenario (S) | S1 | S2 | S3 | S4 |
|---|---|---|---|---|
| Maximum processing and analysis time, $T_{pa}$ (s) | 5.3 | 10.6 | 15.9 | 21.2 |
| Transmission time to the processing server, $T_t$ (s) | 234.7 | 229.4 | 224.1 | 218.8 |
| Data rate to transmit ECG signal to processing server (kbps) | 8.181 | 8.369 | 8.567 | 8.775 |
| Data rate to transmit analysed ECG signal to cloud storage (kbps) | 4.688 | 2.344 | 1.563 | 1.172 |
| Transmission time to the cloud storage (s) | 27.03 | 54.07 | 81.1 | 108.13 |

The power profile considered in this work for all networking equipment and processing server consists of idle power and a linear load proportional power. The idle power of network equipment and processing server is 90% [18], [19] and 54% [20], respectively of the power consumption at maximum utilisation. The networking devices are shared by multiple applications while the considered processing servers are dedicated for the healthcare application. As discussed above, the healthcare application is considered to contribute to 0.3% of the idle power of the networking devices. Table 2 shows the maximum power consumption and capacity of the devices used in the evaluated architecture.

*Table 2. Power Consumption and Capacity of devices.*

| Type of device | Power (Watts) | Capacity | Type of device | Power (Watt) | Capacity |
|---|---|---|---|---|---|
| Access point (Wi-Fi) [21] | 21 | 0.3 Gbps | Cloud Switch [19] | 2020 | 320 Gbps |
| ONU [22] | 8 | 3.75 Gbps | Cloud Storage [23] | 4900 | 75.6TB |
| OLT [24] | 20 | 128 Gbps | Core router [19] | 12300 | 4480 Gbps |
| Aggregation switch [19] | 1766 | 256 Gbps | Content server [23] | 380.8 | 1.8 Gbps |
| Processing server (PandaBoard) [25] | 3.96 | - | Aggregation router/Cloud router [19] | 4550 | 560 Gbps |

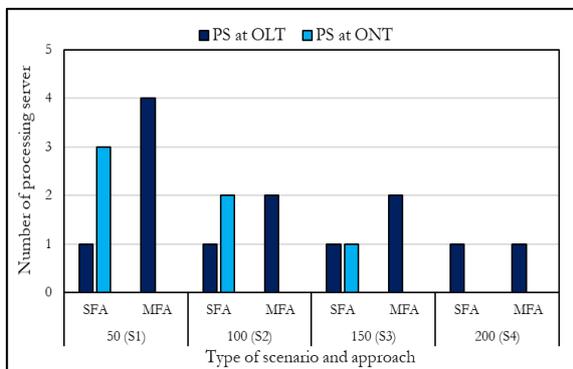 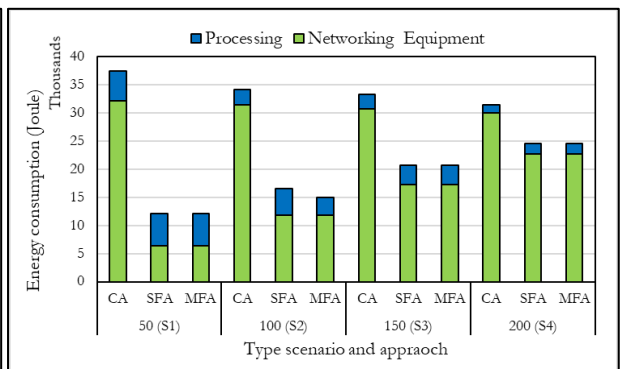

*Figure 2. Processing server placement in the network*  *Figure 3. Energy consumption of network and processing*

Figure 2 shows the optimum location of the processing servers for SFA and MFA under the four scenarios. In SFA, the OLT is always chosen to place one processing server in all scenarios. However, due to the limitation of placing only one processing server at the OLT, the ONTs are chosen to place the remaining processing servers to support the demand for processing and analysis of the ECG signals from patients. The number of ONTs chosen to place processing servers is reduced with increase in the number of maximum patients that can be served by a processing server. Meanwhile, allowing more than one processing server to be placed at the OLT in MFA results in placing all the processing servers at the OLT. This is because, the OLT is the nearest shared point to the patients, (the OLT is connected to all access points in the network). This also reduces the number of hops to transmit the ECG signal to the processing server compared to locating the processing servers at the ONT since the processing servers are shared by multiple patients in different access points. Furthermore, in scenario 2, where $Pat_{max}$ is 100, the SFA utilised three processing servers while MFA utilised only two. This is due to the link capacity constraint between the ONT and OLT. The link has to send the ECG signal of the remaining 100 patients to the same processing server at the ONT while the other 100 patients are served by the processing server at the OLT.

Figure 3 shows the energy consumption of both processing and networking equipment under the four scenarios for SFA, MFA and the conventional approach (CA) where in the CA the processing is done fully in the cloud. The results indicate that a total energy saving of 68% can be achieved when processing the data at the fog for both approaches compared to processing at the central cloud when $Pat_{max}$ is 50. This is because the location of the processing server at fog is near to the patient compared to the central cloud, hence the energy consumed by the networking equipment to transport the raw ECG signal for processing is reduced. However, due to the limitation of link capacity at the network edge to transport the analysed ECG signal to the central cloud, the total energy saving is reduced to 22% when the capability of the processing server serving patients increases to 200 patients. This is because increasing the number of patients served at a processing server reduces the data rate to transmit the analysed data as the link is shared by more patients. This increases the utilisation time of the networking equipment to transport the analysed ECG signal, hence increases the total energy consumption in SFA and MFA. The energy consumption in SFA and MFA in scenario 1 and 3 are almost similar. This is because, the power consumption of the ONT and OLT are approximately 100x smaller compared to the network equipment in metro and core network, which results in 0.2% and 0.1% energy saving in MFA compared to SFA under scenarios 1 and 3, respectively. Meanwhile, in scenario 2, the energy saving in MFA compared to SFA is 9%, mainly due to the high number of processing servers used in SFA compared to MFA. However, the energy consumption of SFA and MFA in scenario 4 are equal as the location and number of processing servers used are the same.

## 4. CONCLUSIONS

In this paper, we optimised the placement of processing servers to process and analyse the ECG signal from patients at the network edge for health monitoring applications. The results show a total energy saving of 68% when performing the processing and analysis of the ECG signal at the network edge compared to processing at the central cloud. However, the energy saving is reduced when the processing server can serve a higher number of patient. This is due to the link capacities limitation at the network edge which results in higher time utilisation of the devices to transport the analysed ECG signal for permanent storage at the central cloud. Besides, the results also indicate that the optimal location to place the processing server in the GPON network is at the OLT as it is the nearest shared point to the patients.

**ACKNOWLEDGEMENT**

The authors would like to acknowledge funding from the Engineering and Physical Sciences Research Council (EPSRC), through INTERNET (EP/H040536/1) and STAR (EP/K016873/1) projects. The first author would like to thank the Ministry of High Education of Malaysia and Universiti Teknikal Malaysia Melaka (UTeM) for funding her PhD scholarship. All data are provided in full in the results section of this paper.

**REFERENCES**
[1]   T. N. Gia, M. Jiang, A. M. Rahmani, T. Westerlund, P. Liljeberg, and H. Tenhunen, "Fog computing in healthcare Internet of Things: A case study on ECG feature extraction," in *IEEE International Conference on Computer and Information Technology; Ubiquitous Computing and Communications; Dependable, Autonomic and Secure Computing; Pervasive Intelligence and Computing*, 2015, pp. 356–363.
[2]   J. M. H. Elmirghani *et al.*, "GreenTouch GreenMeter Core Network Energy-Efficiency Improvement Measures and Optimization," *J. Opt. Commun. Netw.*, vol. 10, no. 2, p. A250, 2018.
[3]   L. Nonde, T. E. H. El-gorashi, and J. M. H. Elmirghani, "Energy Efficient Virtual Network Embedding for Cloud Networks Leonard," *J. Light. Technol.*, vol. 33, no. 9, pp. 1828–1849, 2015.
[4]   A. Q. Lawey, T. E. H. El-Gorashi, and J. M. H. Elmirghani, "BitTorrent Content Distribution in Optical Networks," *J. Light. Technol.*, vol. 32, no. 21, pp. 4209–4225, 2014.
[5]   N. I. Osman, T. El-Gorashi, L. Krug, and J. M. H. Elmirghani, "Energy-efficient future high-definition TV," *J. Light. Technol.*, vol. 32, no. 13, pp. 2364–2381, 2014.
[6]   X. Dong, T. El-gorashi, and J. M. H. Elmirghani, "On the energy efficiency of physical topology design

for IP over WDM networks," *J. Light. Technol.*, vol. 30, no. 12, pp. 1931–1942, 2012.
[7] X. Dong, T. El-Gorashi, and J. M. H. Elmirghani, "Green IP over WDM networks with data centers," *J. Light. Technol.*, vol. 29, no. 12, pp. 1861–1880, 2011.
[8] X. Dong, T. El-Gorashi, and J. M. H. Elmirghani, "IP Over WDM Networks Employing Renewable Energy Sources," *J. Light. Technol.*, vol. 29, no. 1, pp. 3–14, 2011.
[9] M. Musa, T. Elgorashi, and J. Elmirghani, "Energy efficient survivable IP-Over-WDM networks with network coding," *J. Opt. Commun. Netw.*, vol. 9, no. 3, pp. 207–217, 2017.
[10] A. M. Al-Salim, A. Q. Lawey, T. E. H. El-Gorashi, and J. M. H. Elmirghani, "Energy Efficient Big Data Networks: Impact of Volume and Variety," *IEEE Trans. Netw. Serv. Manag.*, vol. 15, no. 1, pp. 458–474, 2018.
[11] P. L. and T. S. I. Azimi, A. Anzanpour, A. M. Rahmani, "Medical Warning System Based on Internet of Things Using Fog Computing," in *2016 International Workshop on Big Data and Information Security (IWBIS)*, 2016, pp. 19–24.
[12] A. M. Rahmani *et al.*, "Exploiting smart e-Health gateways at the edge of healthcare Internet-of-Things: A fog computing approach," *Futur. Gener. Comput. Syst.*, vol. 78, no. Part 2, pp. 641–658, 2018.
[13] P. Verma and S. K. Sood, "Fog Assisted-IoT Enabled Patient Health Monitoring in Smart Homes," *IEEE Internet Things J.*, pp. 1–8, 2018.
[14] I. Cale, A. Salihovic, and M. Ivekovic, "GPON ( Gigabit Passive Optical Network )," in *29th International Conference on Information Technology Interfaces*, 2007, pp. 679–684.
[15] P. Kakria, N. K. Tripathi, and P. Kitipawang, "A real-time health monitoring system for remote cardiac patients using smartphone and wearable sensors," *Int. J. Telemed. Appl.*, vol. 2015, pp. 1–11, 2015.
[16] Cisco, "The Zettabyte Era : Trends and Analysis," 2017.
[17] R. Prieto, "Cisco Visual Networking Index Predicts Near-Tripling of IP Traffic by 2020," 2016.
[18] T. A. and R. S. T. F. Jalali, K. Hinton, R. Ayre, "Fog Computing May Help to Save Energy in Cloud Computing," *IEEE J. Sel. Areas Commun.*, vol. 34, no. 5, pp. 1728–1739, 2016.
[19] R. W. A. A. and R. S. T. A. Vishwanath, F. Jalali, K. Hinton, T. Alpcan, "Energy Consumption Comparison of Interactive Cloud-Based and Local Applications," *IEEE J. Sel. Areas Commun.*, vol. 33, no. 4, pp. 616–626, 2015.
[20] N. Balakrishnan, "Building and benchmarking a low power ARM cluster," 2012.
[21] C. Gray, R. Ayre, K. Hinton, and R. S. Tucker, "Power consumption of IoT access network technologies," in *2015 IEEE International Conference on Communication Workshop (ICCW)*, 2015, pp. 2818–2823.
[22] Alcatel-Lucent, "Alcatel-Lucent 7368 ISAM ONT G-240G-A," 2014.
[23] A. Q. Lawey, T. E. H. El-Gorashi, and J. M. H. Elmirghani, "Distributed energy efficient clouds over core networks," *J. Light. Technol.*, vol. 32, no. 7, pp. 1261–1281, 2014.
[24] Eltex, "GPON Optical Line Terminal Data Sheet," 2015.
[25] B. Pang, "Energy Consumption Analysis of ARM- based System," Aalto University, 2011.